# Thermal Annealing Effect on Electrical and Structural Properties of Tungsten Carbide Schottky Contacts on AlGaN/GaN heterostructures

G. Greco [1], S. Di Franco [1], C. Bongiorno [1], E. Grzanka [2], M. Leszczynski [2],

F. Giannazzo [1] and F. Roccaforte [1]

[1] Consiglio Nazionale delle Ricerche – Istituto per la Microelettronica e Microsistemi (CNR-IMM), Strada VIII, n. 5 – Zona Industriale, 95121 Catania, Italy

[2] Institute of High Pressure Physics - PAS, Sokolowska 29/37, 01-152 Warsaw, Poland

**Abstract**

Tungsten carbide (WC) contacts have been investigated as a novel gold-free Schottky metallization for AlGaN/GaN heterostructures. The evolution of the electrical and structural/compositional properties of the WC/AlGaN contact has been monitored as a function of the annealing temperature in the range from 400 to 800°C. The Schottky barrier height ($\Phi_B$) at WC/AlGaN interface, extracted from the forward current-voltage characteristics of the diode, decreased from ~0.8 eV in the as-deposited and 400°C annealed sample, to 0.56 eV after annealing at 800 °C. This large reduction of $\Phi_B$ was accompanied by a corresponding increase of the reverse bias leakage current. Transmission electron microscopy coupled to electron energy loss spectroscopy analyses revealed the presence of oxygen (O) uniformly distributed in the WC layer, both in the as-deposited and 400°C annealed sample. Conversely, oxygen accumulation in a 2-3 nm thin W-O-C layer at the interface with AlGaN was observed after the annealing at 800 °C, as well as the formation of $W_2C$ grains within the film (confirmed by X-ray diffraction analyses). The formation of this interfacial W-O-C layer is plausibly the main origin of the decreased $\Phi_B$ and the increased leakage current in the 800°C annealed Schottky diode, whereas the decreased O content inside the WC film can explain the reduced resistivity of the metal layer. The results provide an assessment of the processing conditions for the application of WC as Schottky contact for AlGaN/GaN heterostructures.

## 1. Introduction

In the last decades, AlGaN/GaN heterostructures have been receiving a great attention from the semiconductor community. A unique feature of these systems is the carrier confinement at the AlGaN/GaN heterointerface in a two-dimensional electron gas (2DEG) having a high sheet density (~$10^{13}$ cm$^{-2}$) and mobility (~ 2000 cm$^2$/V/s) [1]. This characteristic, combined with the large band gap and high critical electric field of GaN-based materials, allows the fabrication of high electron mobility transistors (HEMTs) operating at high-frequency and high-voltage, suitable for power electronics applications [2]. An interesting aspect is the possibility to grow AlGaN/GaN heterostructures onto large area Si substrates, which makes these systems very promising for the monolithic integration of GaN-based devices with Si technology. In this context, making GaN-devices fabrication fully compatible with Si technology requires the use of Au-free Ohmic and Schottky metallizations, in order to prevent undesired sources of contamination and to reduce the fabrication costs.

In literature, several works discussed the key role of Au in the achievement of a low contact resistance and, hence, the possible routes to form good Au-free Ohmic contacts on AlGaN/GaN heterostructures [3,4,5,6,7,8,9,10].

On the other hand, although Schottky contacts are probably less studied, they are equally important in AlGaN/GaN HEMTs, as they modulate the output current in the device and have a strong impact on the leakage current in the off-state [2]. For that reason, a good understanding of the carrier transport mechanisms in Au-free Schottky contacts and their response to thermal budgets are required [11,12].

Typically, Schottky contacts to GaN-based materials are formed using Ni/Au [13,14] or Pt/Au [15] bilayers. However, many Au-free Schottky metallizations for AlGaN/GaN HEMTs have been proposed as an alternative, e.g., Ti [16], Cr [16], Pt [17], Pd [17,18], Ir [18,19], Mo [20], ITO [21,22], TaN [22,23], TiN [24,25]. These latter (TaN and TiN) exhibited some interesting features, such as a low leakage current combined with a good thermal stability [23,24]. The electrical properties of these systems are strongly influenced by the deposition technique, as their work function can vary in wide range (i.e., 4.13 – 5.05 eV for TaN [26] and 3.8 – 4.6 eV for TiN) [27,28,29].

Recently, Tungsten-based compounds (W [30], WSi [18] or WN [31]), have been proposed as Schottky contacts on AlGaN/GaN heterostructures as they showed promising results in terms of ON/OFF current ratio and turn-on voltage uniformity. Among W-based compounds, Tungsten Carbide (WC) is known for its high mechanical and wear resistance properties [32,33]. In recent years the use of WC has also been considered to obtain stable Schottky contacts on silicon carbide [34,35] and diamond [36,37]. However its possible use on GaN-based materials and heterostructures has not been investigated yet.

In this paper, the electrical behavior of Tungsten Carbide (WC) Schottky contacts on AlGaN/GaN heterostructures has been studied and correlated with the structural and compositional evolution of the WC/AlGaN system after thermal annealing. This study revealed good electrical performances of the diodes up an annealing temperature of 600°C and a degradation of the forward and reverse bias characteristics after 800 °C, which was explained by the reduction of the WC/AlGaN Schottky barrier height associated to the formation of a thin oxygen-rich W-O-C layer at the interface.

## 2. Experimental

AlGaN/GaN heterostructures grown on Si, with an AlGaN thickness of 16 nm and an Al content of 26%, have been used in this work. The fabrication of Schottky Barrier Diodes started with the definition of the Ohmic contact by optical lithography and lift-off technique. Non-recessed Au-free Ohmic contact have been formed by Ti/Al/Ti multilayer annealed at 600°C for 3 minutes [38]. Then, Schottky contacts have been obtained by the deposition of 100 nm of WC layer. Such structures consist of an inner circular Schottky contact of radius of 40 μm and an Ohmic contact formed by a circular ring, which surrounds the Schottky electrode. Also the Schottky contacts were defined by optical lithography followed by the lift-off technique. Both Ohmic and Schottky metals have been deposited using a Quorum Q300T D sputter system. Post deposition rapid annealing (60 s) treatments have been performed in Ar atmosphere in the temperature range 400-800°C, by using a Jipelec JetFirst 150 furnace. Atomic force microscopy (AFM) scans have been acquired to monitor the surface morphology of the investigated material using a Dimension 3100 microscope with Nanoscope V controller. The current–voltage (I–V) measurements were performed on a Karl Suss Microtec probe station equipped with a HP 4156B parameter analyser. The structural analyses were conducted by X-ray diffraction (XRD) and Transmission Electron Microscopy (TEM). XRD measurements were carried out using a Malvern Panalytical Empyrean High Resolution X-ray Diffractometer. TEM analyses in cross-section were performed using a 200 kV JEOL 2010 F microscope, equipped with Gatan imaging filter (GIF) spectrometer allowing electron energy loss spectroscopy (EELS).

## 3. Results and discussion

Firstly, the surface morphology of the WC layer deposited onto the AlGaN/GaN heterostructure has been monitored by AFM. Fig.1 shows four representative AFM images acquired on 50×50 µm² scan areas of the bare AlGaN surface (a), of the as-deposited WC contact (b) and after annealing at 600 °C (c) and 800°C (d). Fig.1e summarizes the evolution of the root mean square (RMS) surface roughness of the WC as a function of the annealing temperature. Each value in the plot is the average of the RMS obtained on several AFM scans and the error bar is the standard deviation. The as-deposited WC layer exhibits a similar surface morphology to that of the AlGaN underneath (RMS=5.1 nm ± 1.1 nm). It can be also noticed that the starting RMS value of the as-deposited WC is preserved up to the annealing temperature of 600°C, whereas a further annealing temperature increase to 800°C results in a higher surface roughness (RMS=7.4 nm ± 1.6 nm).

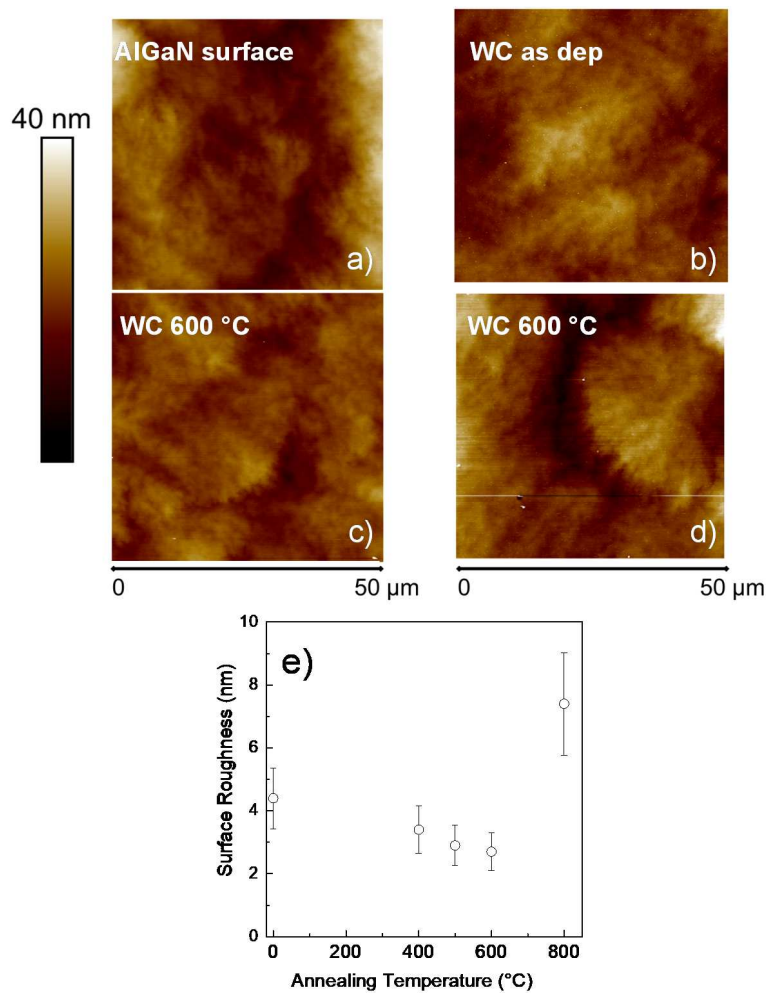

Fig. 1: AFM images of the bare AlGaN surface before metal deposition (a), of the as deposited (b), the 600°C (c) and 800°C annealed (d) WC contacts. Average values of the surface roughness (RMS) of the WC contact, determined by the AFM images, as function of the annealing temperature (e).

The evolution of the resistivity of the WC layer deposited on AlGaN/GaN heterostructures was also monitored by four-points probe measurements as a function of the annealing temperature (see Fig. 2). In particular, a resistivity of about 2×10$^{-4}$ Ωcm was evaluated for the as deposited WC layer. The metal resistivity decreased down to 1×10$^{-4}$ Ωcm after the thermal treatment at 800 °C. Resistivity values in the order of 10$^{-4}$ Ωcm are typical of sputtered WC thin films, as can be seen from the good agreement with literature values measured on WC films on sapphire [36], which is reported in the same graph for comparison.

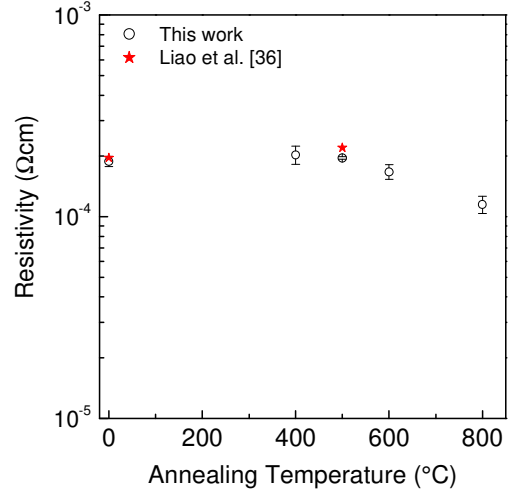

Fig. 2: Resistivity of the WC layer deposited on AlGaN/GaN heterostructure as function of the annealing temperature. The literature data of WC films on sapphire, taken from reference [36], are reported for comparison.

Then, the electrical behavior of the diodes fabricated on AlGaN/GaN heterostructures by using a WC Schottky contact have been investigated. Fig. 3a shows the forward current density-voltage (J-V) characteristics of diodes with the as-deposited WC contact and after different annealing temperatures (up to 800 °C). A decrease of the forward current was initially observed from the as-deposited contact to the one annealed at 400 °C, whereas a higher current injection was found at the annealing temperature of 600 °C and 800 °C. For all samples, the semilog plot of the J-V curves displays a linear increase of the current in the bias range from 0 to ~1 V, followed by a saturation behavior at higher forward bias values (V > 1.0-1.5 V) due to the series resistance contribution. In ideal Schottky barrier diodes, the carrier transport is typically ruled by the thermionic emission (TE) model, which in the linear region can be approximated as:

$$J = J_S \left[ exp\left(\frac{qV}{nkT}\right) - 1 \right] \approx J_S \, exp\left(\frac{qV}{nkT}\right) \qquad \text{Eq. 1}$$

with

$$J_S = A^* T^2 \, exp\left(-\frac{q\Phi_B}{kT}\right) \qquad \text{Eq. 2}$$

where A* is the Richardson constant (32 A/(cm$^2$K$^2$) for AlGaN) [14,39], T is the measurement temperature, k is the Boltzmann constant, and $\Phi_B$ and *n* are the Schottky barrier height and the ideality factor, respectively. From the fit of the experimental curves with the TE model, an ideality factor of 1.59 and a barrier height of 0.82 eV were evaluated for the as-deposited WC Schottky diode. The theoretical barrier height value $\Phi_B$ for an ideal WC/AlGaN interface depends on the WC work function $\Phi_{m(WC)}$ according to the Schottky-Mott relation $\Phi_B = \Phi_{m(WC)} - \chi_{AlGaN}$, where $\chi_{AlGaN}$ is the value of the electron affinity for an AlGaN layer with 26% Al content. For the WC, a work function of 5.2 eV can be estimated [40,41]. In this condition, an ideal barrier height of 1.1 eV is expected, which is much higher than that obtained by the fit of experimental J-V curves. A similar deviation from the ideal barrier value have been observed in Schottky contacts to AlGaN/GaN heterostructures for different metallizations. Despite the physical origin of this deviation is still debated, many works ascribe this behavior to the material quality, the presence of surface states, effects of polarization charges or the presence of the 2DEG [42,43,44,45].

After thermal annealing at 400 °C, a decrease of the ideality factor from 1.59 to 1.50 is observed, accompanied by a slight increase of $\Phi_B$ up to 0.85 eV. By further raising the annealing temperature to 600 °C and 800 °C, the ideality factor significantly increased to 2.13 and 2.65, while the barrier height decreased to 0.72 eV and 0.56 eV (Fig. 3b).

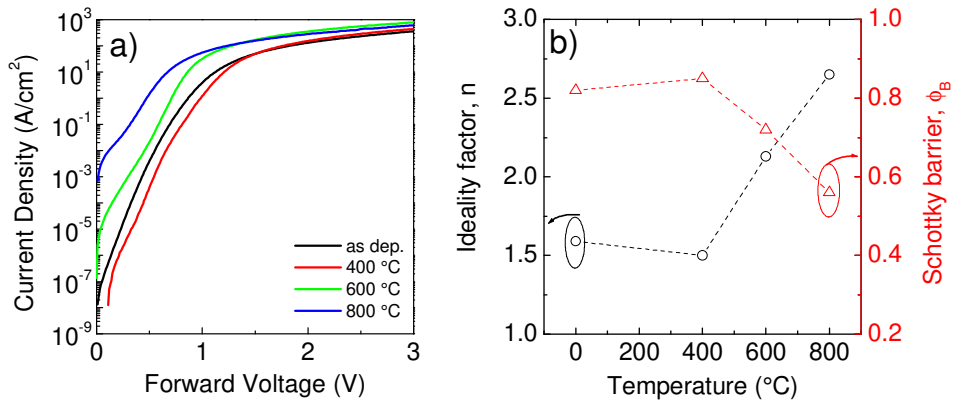

Fig. 3: (a) Forward J-V characteristics of WC Schottky diodes subjected to annealing processes at different temperatures; (b) Ideality factor and Schottky barrier height as a function of the annealing temperature.

The current-voltage characteristics of the WC/AlGaN/GaN Schottky diodes under negative (reverse) bias polarization have been also investigated. Fig. 4a shows the reverse J-V curves acquired after different annealing temperatures. All the acquired characteristics exhibited a first region at lower negative bias where the current increases with the voltage, followed by a second region at higher bias with constant current (i.e., independent of the bias), corresponding to the complete depletion of the 2DEG. Evidently, the annealing treatments have a notable impact on the reverse characteristics of the diodes. In particular, the leakage current density at a reverse bias V$_R$=-10 V firstly decreased from 4.1×10$^{-4}$ A/cm$^2$ in the as-deposited WC contact to 2.2×10$^{-5}$ A/cm$^2$ after annealing at 400 °C, while it increased again to 1.3×10$^{-2}$ A/cm$^2$ in the sample annealed at higher temperature of 600 °C. A further increase of the leakage current density to 3.3 A/cm$^2$ (at V$_R$=-10 V) was observed after the contact was annealed at 800 °C (Fig. 4b). The increase of the reverse current can be justified by the reduction observed in the Schottky barrier height.

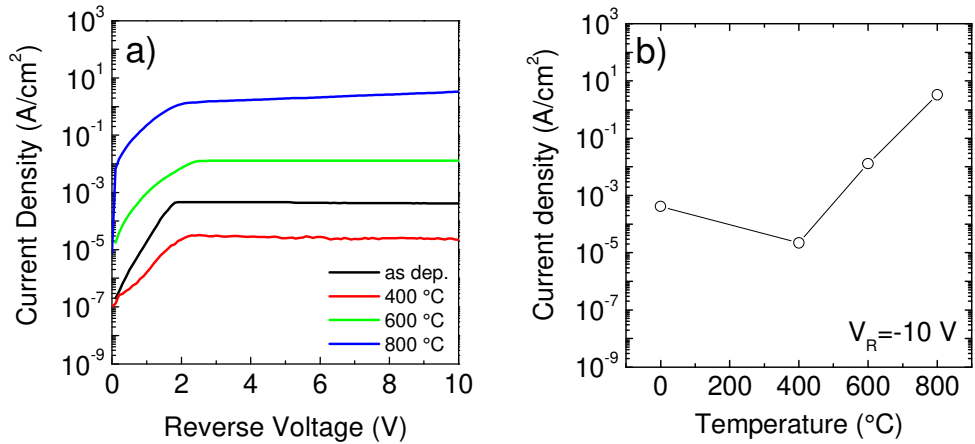

Fig. 4: (a) Reverse J-V characteristics of WC Schottky diodes subjected to annealing processes at different temperatures. (b) Reverse leakage current density at $V_R$=-10V as a function of the annealing temperature.

In order to understand the origin of the electrical behavior of the Schottky contacts, a structural investigation of the WC layer has been carried out by means XRD and TEM analyses.

In particular, no significant structural changes of the WC layer were shown by XRD analysis (see Fig. 5) in the as-deposited contact and after the thermal annealing at 400°C. On the other hand, by increasing the annealing temperature to 800 °C, some peaks associated to the presence of $W_2C$ appear in the XRD spectrum, despite the WC remains the dominant phase in the metal layer. As a matter of fact, some literature works report on the formation of $W_2C$ phase after thermal annealing [46] and, in general, on the possible formation of other W-C phases at temperatures higher than 500 °C [47].

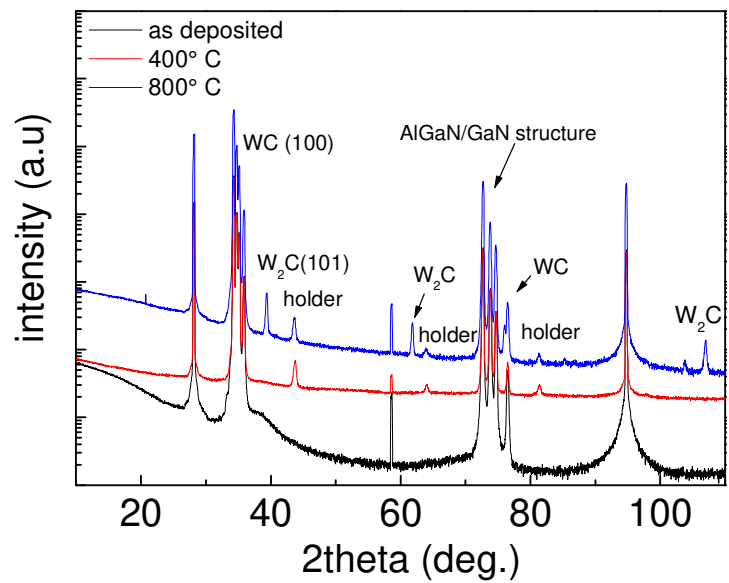

Fig. 5: XRD patterns of WC/AlGaN/GaN system after metal deposition (as deposited) and after annealing at 400 °C and 800 °C.

   In Fig. 6 cross-sectional bright field TEM micrographs of the WC/AlGaN interface are shown for the as deposited and annealed samples at 400°C and 800 °C. No significant differences in the WC film and in the WC/AlGaN interface can be observed from the TEM analyses on the as-deposited (see Fig. 6a) and on the 400 °C annealed sample (see Fig. 6b), Instead, after annealing at a temperature of 800 °C, a 2-3 nm thick interface layer appears between the AlGaN and the WC layer (see Fig. 6c)). EELS analysis confirmed the presence of both W, C and O in this thin layer. Moreover, at this annealing temperature the presence of W-rich grains was detected in the metal film, which is likely associated with the $W_2C$ phase detected by XRD analysis. In order to investigate the origin of the oxygen incorporated at the interface layer, EELS analyses have been carried out also to monitor the oxygen content within the as-deposited and annealed WC metal films. Fig. 6d displays three typical EELS spectra collected within the metal film in the as-deposited sample and after annealing at 400°C and 800°C. Noteworthy, the characteristic oxygen-related edge at 532 eV is clearly visible in the as deposited and 400 °C annealed WC layer, thus indicating the presence of oxygen distributed within the film. Oxygen incorporation into the WC layer represents a long standing problem in deposited WC layer [48]. Instead, the oxygen edge could not be detected in the EELS spectra collected in the WC layer annealed at 800°C. A possible explanation can be found in the diffusion of the oxygen toward the surface and AlGaN interfaces during annealing treatment at very high temperature (800 °C). In this scenario, the oxygen reaching the AlGaN interface leads to the formation of the oxygen rich W-O-C layer detected by TEM.  Fig. 7 displays a schematic representation of the evolution of system, to explain the correlation between the electrical behaviour with the structural and compositional properties of the WC/AlGaN contact during annealing. In particular the degradation of the current-voltage characteristics in forward and reverse bias can be correlated to the modification of the WC/AlGaN interface after annealing at high temperature (800 °C), with the formation of a W-O-C containing layer. The presence of $W_2C$ grains detected inside the films, which are far from the interface, plays probably no role on the electrical characteristics of the diodes. On the other hand, the oxygen desorption from the WC layer can explain the decrease of the film resistivity observed upon annealing.

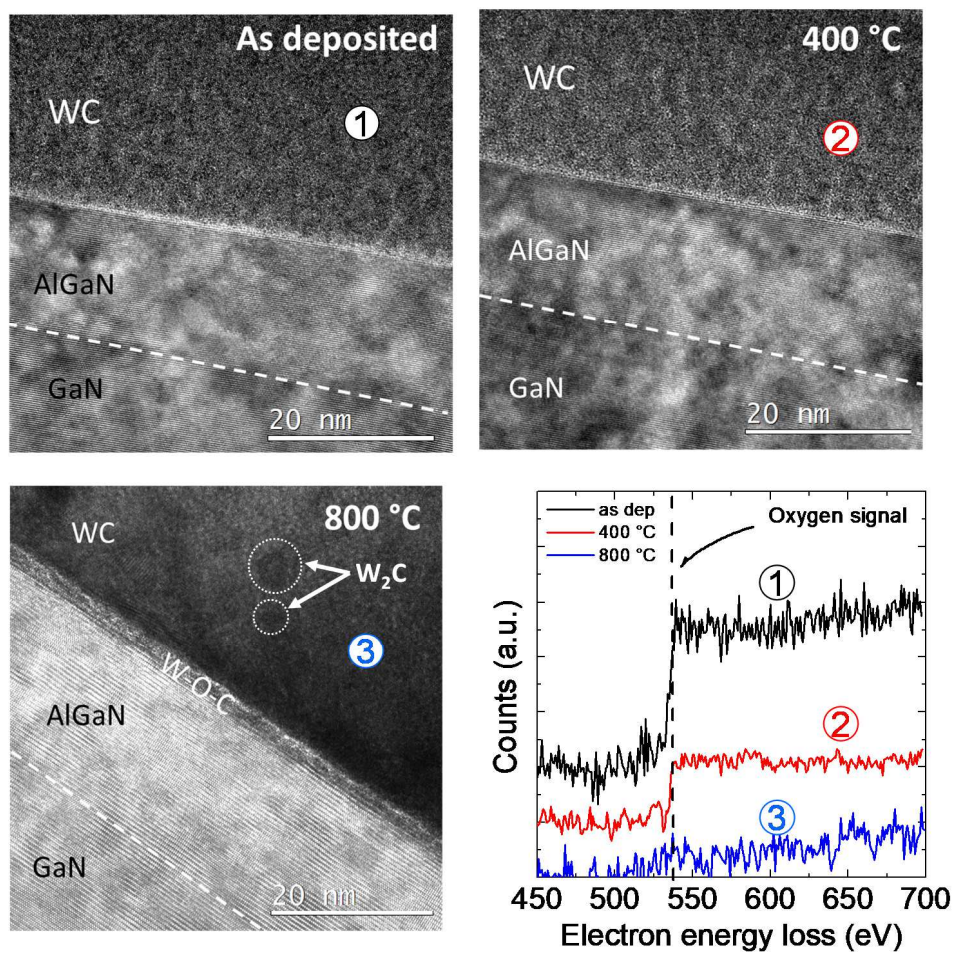

Fig. 6: Bright field TEM micrographs in cross section for WC/AlGaN/GaN contacts (a) as deposited and annealed at temperature of (b) 400 °C and (c) 800 °C. (d) The electron energy loss spectra related to the signal of oxygen acquired in the regions 1, 2 and 3 of the as deposited and annealed WC layer.

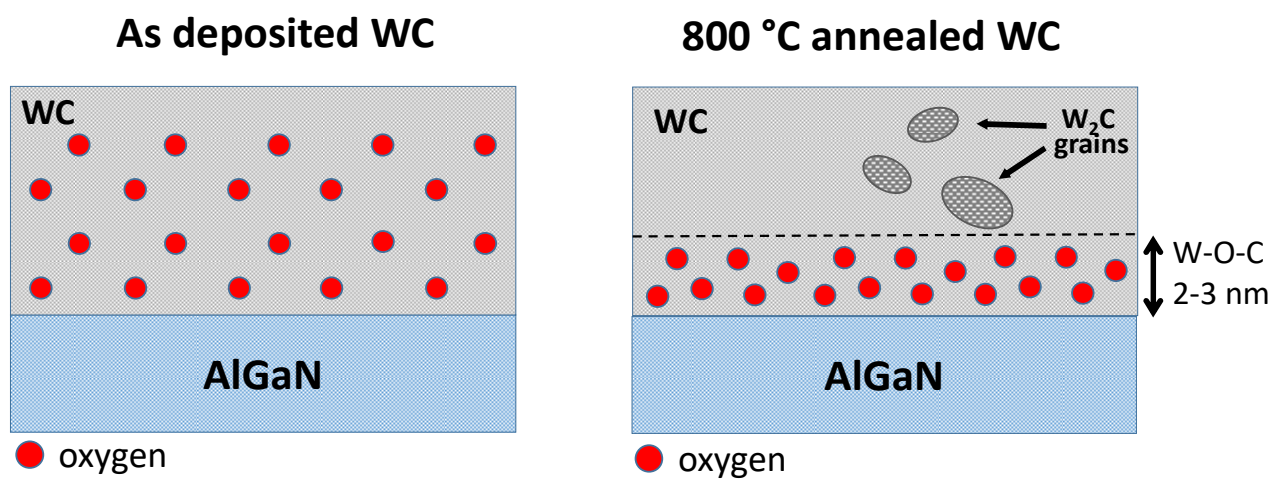

Fig. 7: Schematic representation of the evolution of the WC layer during annealing at 800°C.

## 4. Conclusion

In conclusion, the evolution of the electrical and structural properties of WC Schottky contacts on AlGaN/GaN heterostructures has been investigated by cross correlating different analyses. The WC layer exhibited a smooth morphology even after annealing at 600°C, which increased at higher temperatures (800°C). The WC/AlGaN Schottky barrier height, determined from the forward characteristics of the diodes, decreased from 0.82 eV (as-deposited) to 0.56 eV (annealed at 800 °C), thus being accompanied by an increase of the reverse leakage current. A microstructural and chemical analysis revealed the presence of oxygen in the WC layer, which accumulates in a 2-3 nm thin W-O-C layer at the interface with AlGaN after the annealing at 800 °C. Moreover, XRD showed the formation of $W_2C$ grains within the film. This scenario can be the main origin of the decreased barrier height and the increased leakage current at 800°C annealed Schottky diode. These results provide useful insights for the practical application of this Au-free metallization as Schottky contact for AlGaN/GaN heterostructures.

## 5. Acknowledgments

This work was performed in the framework of the Italian National project PON ARS01_01007 EleGaNTe (*"Electronics on GaN based Technologies"*). The authors would like also to acknowledge the bilateral project GaNIMEDe (Gallium Nitride Innovative Micro-Electronics DEvices) within the Executive Programme for Scientific and Technological Cooperation between The Italian Republic and the Republic of Poland for the years 2019-2020 and the bilateral project ETNA *("Energy efficiency Through Novel AlGaN/GaN heterostructures")* within the CNR-PAS Cooperation Agreement for the years 2020-2021.